%% file: constellation-shaping-paper.tex
\documentclass[conference]{IEEEtran}
\IEEEoverridecommandlockouts
\usepackage[T1]{fontenc}
\usepackage[utf8]{inputenc}
\usepackage{url}
\usepackage{cite}
\usepackage{amsmath,amssymb,amsfonts}
\usepackage{algorithmic}
\usepackage{graphicx}
\usepackage{textcomp}
\usepackage{xcolor}
\usepackage{booktabs}
\usepackage[nolist]{acronym}
\usepackage{siunitx}
\usepackage{pgfplots}
\usepackage{soul}
\pgfplotsset{compat=1.17}

\usepgfplotslibrary{groupplots}
\usetikzlibrary{arrows.meta,positioning,calc,fit,shapes.arrows}
\graphicspath{{figs/}}

\definecolor{clr1}{RGB}{0,114,189}
\definecolor{clr2}{RGB}{217,83,25}
\definecolor{clr3}{RGB}{126,47,142}
\definecolor{clr4}{RGB}{119,172,48}
\definecolor{clr5}{RGB}{162,20,47}
\definecolor{clr6}{RGB}{77,190,238}

\newcommand{\dftsofdm}{\ac{dft}\nobreakdash-s\nobreakdash-\ac{ofdm}}

\pgfplotsset{
  resultaxis/.style={
    width=0.95\linewidth,
    tick label style={font=\footnotesize},
    label style={font=\small},
    legend style={font=\scriptsize},
    legend cell align={left},
    grid=both,
    major grid style={solid, gray!18},
    minor grid style={dotted, gray!18},
    height=6cm,
    every axis plot/.append style={line width=1.2pt},
    mark size=1.8pt,
  },
}
\def\BibTeX{{\rm B\kern-.05em{\sc i\kern-.025em b}\kern-.08em
    T\kern-.1667em\lower.7ex\hbox{E}\kern-.125emX}}

\begin{document}
\input{acronyms}

\title{Net-Gain-Optimized Constellation Shaping for the 5G NR DFT-s-OFDM Uplink:\\
Probabilistic, Geometric, and Joint Designs}

\author{\IEEEauthorblockN{G\"unther Lindorfer}
\IEEEauthorblockA{\textit{Institute for Communications Engineering and RF-Systems} \\
\textit{Johannes Kepler University Linz}\\
Linz, Austria \\
guenther.lindorfer@jku.at}
\and
\IEEEauthorblockN{Andreas Springer}
\IEEEauthorblockA{\textit{Institute for Communications Engineering and RF-Systems} \\
\textit{Johannes Kepler University Linz}\\
Linz, Austria \\
andreas.springer@jku.at}
}

\maketitle

\begin{abstract}
\Acl{cs} improves the \acl{air} of coded modulation. On the single-carrier
uplink waveform of 5G \acl{nr}, \ac{dft}-spread \ac{ofdm}, the same shaping can also raise the
\ac{papr}. A higher \ac{papr} forces more \acl{pa} back-off and reduces the link budget.
Prior work characterized this penalty for fixed probabilistically and geometrically shaped constellations, but
did not turn the trade-off into a design objective. We optimize \ac{cs} to
maximize the net-gain, i.e., the sum of the communication gain and the \ac{papr} gain, at the fixed spectral efficiency of \num{4}~bits per channel use for 64-QAM. We apply it to
\ac{ps}, \ac{gs}, and a joint scheme that combines both. As the coded block error rate cannot be evaluated inside a gradient loop, we use the \acl{air} as a surrogate for the communication
performance and the \ac{papr} surrogate as differentiable cost functions for the optimizer. The learned constellations are then validated on a complete \dftsofdm{} link-level chain.
Conventional Maxwell--Boltzmann \ac{ps} yields a negative net-gain, but
net-gain-aware optimization removes this deficit in every method. \ac{gs} converts
it into a substantial gain, as it
not only improves the communication performance but also lowers the \ac{papr} below the uniform
64-QAM reference, achieving the largest net-gain among the \ac{cs} methods studied.
\end{abstract}
\acresetall


\section{INTRODUCTION}
In 5G \ac{nr} the \ac{dl} employs \ac{cp}-\ac{ofdm}, whereas the \ac{ul} additionally
offers \ac{dft}-spread \ac{ofdm} (\dftsofdm{}), also known as \ac{sc-fdma}~\cite{myung2006singlecarrier}. As a multi-carrier waveform, \ac{cp}-\ac{ofdm} exhibits a high
\ac{papr}, so the \ac{pa} must be operated with a large \ac{obo} to remain in its linear region
and avoid spectral regrowth and in-band
distortion~\cite{rapp1991effects,rahmatallah2013papr}.
This back-off lowers the mean radiated power the \ac{pa} can deliver within its linearity limits, and hence the link budget, a penalty that is most severe on the power-limited \ac{ul}. Since \dftsofdm{} is a single-carrier
waveform with substantially lower \ac{papr}, it is the preferred \ac{ul} waveform for power-limited
transmission. This benefit is usually associated with the lowest modulation orders, but
\dftsofdm{} also supports high-order constellations, up to a maximum of
256-QAM~\cite{3gpp-38.211}. Thus, any optimization technique applied to the waveform must be assessed not only
by its communication performance but also by its effect on \ac{papr}.

\Ac{cs} is a mature and effective way to close the gap between uniform \ac{qam}
and the capacity~\cite{6773024} of the \ac{awgn} channel. This gap is up to about
\SI{1.53}{\deci\bel}~\cite{8640810,256499}. Two complementary \ac{cs} mechanisms exist. In
\ac{ps} the constellation points remain on a regular grid but are transmitted with non-uniform
probabilities, classically the \ac{mb} distribution that is entropy-optimal under
an average-power constraint~\cite{256499,8640810}.
In \ac{gs} the probabilities stay uniform while the point positions are relocated to a
non-uniform layout~\cite{7419222}. The two mechanisms can also be combined into a
joint design that shapes positions and probabilities simultaneously~\cite{9024567}.

The gains of \ac{cs} are usually reported as an information-rate improvement on the \ac{awgn}
channel, a metric that depends on the transmitted symbols but not on the waveform that
carries them. On multi-carrier \ac{cp}-\ac{ofdm}, \ac{cs}
barely affects the \ac{papr}~\cite{8227810}, however, for \dftsofdm{} this is no
longer true. All \ac{cs} methods presented in this paper raise the \ac{papr} when
optimized for communication performance significantly. On the other hand, if the \ac{papr} is optimized, the \ac{air} for a given \ac{snr} is reduced.
Thus, the \ac{papr} and the \ac{air} are competing objectives, and the net effect of shaping on the link budget is not obvious.
We turn that trade-off into an explicit design objective, the \emph{net-gain}, i.e., the sum of
the communication gain and the \ac{papr} gain. A gradient-based framework then optimizes \ac{ps}, \ac{gs}, and joint shaping against this objective on the \dftsofdm{} \ac{ul} at a fixed \ac{seff}.

The remainder of the paper is organized as follows.
Section~\ref{sec:shaping} reviews the three \ac{cs} methods and the \ac{air}.
Section~\ref{sec:system} presents the system model for the \dftsofdm{} \ac{ul} transmitter and receiver, including the \ac{pas} architecture. Section~\ref{sec:framework} defines the net-gain and the
optimization framework. Section~\ref{sec:results} describes the
link-level simulation and presents the results for all considered schemes. Section~\ref{sec:conclusion}
concludes the paper.

\emph{Notation:} An uppercase letter $X$ denotes a random variable, a lowercase $x$ a realization,
a calligraphic letter $\mathcal{X}$ a set, and $|\mathcal{X}|$ is its cardinality. $P_X(\cdot)$ is
the probability mass function of $X$ and $\Pr(\cdot)$ a probability.

\section{CONSTELLATION SHAPING}
\label{sec:shaping}
Let $X$ be the channel input drawn from a constellation $\mathcal{X}=\{x_1,\dots,x_M\}$ with
probabilities $P_X(x_i)$. \Ac{cs} adjusts the symbol probabilities $P_X$, the point
positions $\mathcal{X}$, or both. It is classically used to maximize the \ac{air}, which
drives the input toward a Gaussian-like distribution~\cite{256499,8640810}. The same degrees of freedom, however, can be directed
at any objective, for instance the \ac{air} alone, the \ac{papr} alone, or a combined metric. In
this work we direct them at the \emph{net-gain}, which weighs communication gain against
\ac{papr}, so the optimized constellations are, by design, not Gaussian-like but
strike a balance between \ac{air} and \ac{papr}.

\subsection{Geometric Shaping}
\Ac{gs} keeps the probabilities uniform and relocates the point positions $\mathcal{X}$ to a \ac{nuc}~\cite{7419222}.
A \ac{nuc} may be \ac{1d}, where the two quadrature components share
the same non-uniform \ac{pam} alphabet and demapping stays per dimension. Alternatively, it may
be \ac{2d}, where the $M$ points are optimized jointly in the plane for a higher gain at the
cost of non-separable demapping~\cite{7419222}. We optimize a \ac{2d} \ac{nuc} for \ac{gs} and use the
ATSC~3.0 64-\ac{nuc}~\cite{7419222}, itself a \ac{2d} \ac{nuc}, as the literature baseline.

\subsection{Probabilistic Shaping}
\Ac{ps} fixes $\mathcal{X}$ to a uniform \ac{qam} grid and shapes the probabilities $P_X$. Under
an average-power constraint the entropy-maximizing distribution is the \ac{mb}
distribution~\cite{256499}
\begin{equation}
P_X(x_i)\;=\;\frac{e^{-\nu\,|x_i|^2}}{\sum_{j=1}^{M} e^{-\nu\,|x_j|^2}},
\label{eq:mb}
\end{equation}
parameterized by a single $\nu\ge 0$ that sets its sharpness: a larger $\nu$ concentrates
probability mass on the low-energy inner points and lowers the entropy $H(X)$, so the target entropy is set by the choice of $\nu$. Setting $\nu=0$ gives the uniform input. We retain this \ac{mb}
baseline for \ac{ps}, realized with the \ac{pas} architecture of Section~\ref{sec:system}.

\subsection{Joint Shaping}
Joint shaping optimizes the point positions and the probabilities
together~\cite{9024567}, combining the two mechanisms within the same \ac{pas}
architecture.

\subsection{Achievable Information Rate}
The communication performance is predicted by the \ac{air} of the coded-modulation
receiver, which decodes the $m=\log_2 M$ bits $B_1,\dots,B_m$ of each 64-QAM symbol by \ac{bmd}.
The demapper computes a separate \ac{llr} for
each bit position, and the binary decoder uses only these \acp{llr}~\cite{7307154}. The \ac{air} of this receiver is the
\ac{bmd} rate~\cite{7307154},~\cite{7523336}
\begin{equation}
R_{\mathrm{BMD}} \;=\; H(\mathbf{B}) \;-\; \sum_{i=1}^{m} H(B_i \mid Y),
\label{eq:gmi}
\end{equation}
where $Y$ is the channel output, \mbox{$\mathbf{B}=(B_1,\dots,B_m)$} the $m$ bits of the transmitted symbol, and \mbox{$\sum_i H(B_i\mid Y)$}
the uncertainty remaining after observing $Y$. Because \ac{pas} shapes the probabilities of the amplitude
bits, they are statistically dependent, so~\eqref{eq:gmi} stays strictly below the
\ac{gmi} $\sum_i I(B_i;Y)$ of independent amplitude bits~\cite{8240991}. For \ac{gs}, whose probabilities stay uniform, the bits are independent and $R_{\mathrm{BMD}}$ and \ac{gmi} coincide. $R_{\mathrm{BMD}}$ is bounded above by the
symbol-wise \ac{mi} $I(X;Y)$~\cite{7523336}, the rate an ideal symbol decoder would achieve. We
optimize $R_{\mathrm{BMD}}$ because it is the rate our bit-metric receiver actually can achieve.

We estimate $R_{\mathrm{BMD}}$ by \ac{mc} averaging of the bit \acp{llr} over the \ac{awgn}
channel~\cite{7523336},
\begin{align}
\widehat{R}_{\mathrm{BMD}} ={}& \frac{1}{N}\sum_{k=1}^{N}\big[-\log_2 P_X(x_k)\big] \nonumber\\
&- \frac{1}{N}\sum_{k=1}^{N}\sum_{i=1}^{m}
\log_2\!\Big(1 + e^{(-1)^{b_{k,i}}\,\Lambda_{k,i}}\Big),
\label{eq:gmi-mc}
\end{align}
where the $N$ samples are $y_k=x_k+w_k$ with $x_k$ drawn from $P_X$ and $w_k$ is \ac{awgn} of variance
$\sigma^2$ set by the \ac{snr}, $b_{k,i}$ is the bit at position $i$ of $x_k$, and the bit
\ac{llr} \cite{7523336} is
\begin{equation}
\Lambda_{k,i} = \log\frac{\sum_{x\in\mathcal{X}_i^{1}} P_X(x)\,e^{-|y_k-x|^2/\sigma^2}}
                       {\sum_{x\in\mathcal{X}_i^{0}} P_X(x)\,e^{-|y_k-x|^2/\sigma^2}},
\label{eq:llr}
\end{equation}
with $\mathcal{X}_i^{b}$ the set of constellation points whose bit at position $i$ equals $b$. The first
term of~\eqref{eq:gmi-mc} is the \ac{mc} average of $-\log_2 P_X(x_k)$, which is an unbiased
estimator of the entropy $H(X)$. This approximation~\eqref{eq:gmi-mc} makes no
assumption on the constellation geometry and is differentiable in
both the positions and the probabilities.

\section{SYSTEM MODEL}
\label{sec:system}

\subsection{DFT-s-OFDM Uplink Transceiver}
\begin{figure*}[t]
\centering
\resizebox{\textwidth}{!}{%
\begin{tikzpicture}[
  font=\footnotesize, >=Latex, node distance=6mm and 9mm,
  box/.style={draw, rounded corners=1.5pt, minimum height=8mm, minimum width=13mm, align=center, inner sep=3pt},
  thrw/.style={circle, draw, fill=white, inner sep=0pt, minimum size=2.4pt},
  vec/.style={-{Implies}, double distance=1.8pt, line width=0.5pt},
]
\node[box] (dm)   at (1.4, 1.2)  {DM};
\node[box] (fec)  at (3.5, 1.2)  {systematic\\\ac{fec}};
\node[box] (pas)  at (5.7, 1.2)  {PAS\\mapper};
\node[box] (fecg) at (3.5,-1.35)  {systematic\\\ac{fec}};
\node[box] (nuc)  at (5.7,-1.35)  {NUC\\mapper};
\node[box] (sp)   at (8.8,-0.15)  {S/P};
\node[box] (dft)  at (10.9,-0.15) {$N$-point\\DFT};
\node[box, right=of dft] (sc)   {subcarrier\\mapping};
\node[box, right=of sc]  (ifft) {IFFT $+$ CP};
\coordinate (isw) at (-0.20,-0.15);
\draw (isw) -- ++(-0.75,0) node[left]{Bit stream};
\fill (isw) circle (1.1pt);
\node[thrw] (it) at (0.45, 0.30) {};
\node[thrw] (ib) at (0.45,-0.55) {};
\draw (isw) -- (it);
\node[font=\scriptsize] at (0.24, 0.50) {PS};
\node[font=\scriptsize] at (0.24,-0.80) {GS};
\draw[->] (it) |- (dm.west);
\draw[->] ($(dm.east)+(0,0.16)$)  -- node[above, pos=0.5]{\scriptsize $\mathrm{I}$} ($(fec.west)+(0,0.16)$);
\draw[->] ($(dm.east)+(0,-0.16)$) -- node[below, pos=0.5]{\scriptsize $\mathrm{Q}$} ($(fec.west)+(0,-0.16)$);
\draw[->] (it) -- (0.45,0.5) -- (3.5,0.5) -- (fec.south);
\draw[->] (fec.east) --  (pas.west);
\draw[->] (ib) |- (fecg.west);
\draw[->] (fecg.east) -- (nuc.west);
\coordinate (osw) at (7.35,-0.15);
\fill (osw) circle (1.1pt);
\node[thrw] (ot) at (6.80, 0.30) {};
\node[thrw] (ob) at (6.80,-0.55) {};
\draw (osw) -- (ot);
\draw[->] (pas.east) -| (ot);
\draw[->] (nuc.east) -| (ob);
\draw[->] (osw) -- node[above]{$d_n$} (sp.west);
\draw[vec] (sp.east)   -- node[above, pos=0.5]{\scriptsize $N$}              (dft.west);
\draw[vec] (dft.east)  -- node[above, pos=0.5]{\scriptsize $N$}              (sc.west);
\draw[vec] (sc.east)   -- node[above, pos=0.5]{\scriptsize $M_{\mathrm{sc}}$}  (ifft.west);
\draw[vec] (ifft.east) -- node[above, pos=0.5]{\scriptsize $J M_{\mathrm{sc}}$} ++(0.95,0) coordinate (vend);
\node[right=0.5mm of vend] {$s_\ell$};
\end{tikzpicture}%
}
\caption{\Ac{dft}-s-\ac{ofdm} \ac{ul} transmitter with \ac{cs}. A two-way switch selects the shaping
method: \ac{ps} and joint shaping use the \ac{dm}, a systematic \ac{fec}, and the \ac{pas}
mapper~\cite{7307154}, whereas \ac{gs} skips the \ac{dm} and maps the \ac{fec}-coded bits directly
onto the \ac{nuc}.}
\label{fig:tx}
\end{figure*}
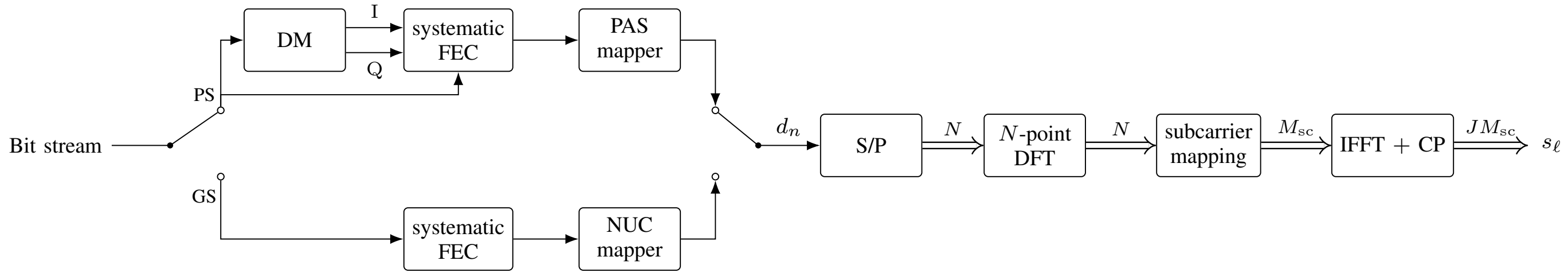

Consider the \dftsofdm{} \ac{ul} transmitter with \ac{cs} of Fig.~\ref{fig:tx}. \Ac{ps} and joint shaping
use a \ac{dm}, a systematic \ac{fec} and the \ac{pas} mapper, whereas \ac{gs} skips the \ac{dm} and
maps the \ac{fec}-coded bits onto the \ac{nuc}. Section~\ref{sec:pas} details how the shaped,
uniform and parity bits are assigned to the amplitude and sign positions of each symbol.

A block of $N$ shaped 64-QAM symbols $d_0,\dots,d_{N-1}$ is first precoded by
an $N$-point \ac{dft} to the frequency-domain sequence~\cite{myung2006singlecarrier}
\begin{equation}
X_k = \frac{1}{\sqrt{N}} \sum_{n=0}^{N-1} d_n\, e^{-\mathrm{j}2\pi kn/N},
\quad k \in \{0,\dots,N-1\}.
\label{eq:dftprecode}
\end{equation}
The $N$
precoded symbols are then mapped onto a contiguous block of subcarriers within an
$M_{\mathrm{sc}}$-point grid, zero-padded, and transformed to the time domain by an \ac{ifft}
oversampled by a factor $J$,
\begin{equation}
s_\ell = \frac{1}{\sqrt{M_{\mathrm{sc}}}} \sum_{k} X'_k\,
e^{\mathrm{j}2\pi k \ell /(J M_{\mathrm{sc}})},
\quad \ell \in \{0,\dots,J M_{\mathrm{sc}}-1\},
\label{eq:ifft}
\end{equation}
where $X'_k$ denotes the subcarrier-mapped and zero-padded spectrum. The oversampling by
$J$\footnote{Without oversampling, discrete-time samples taken at the Nyquist rate can
miss the true peak of the continuous-time waveform, which may occur between two samples. Since
the \ac{papr} and hence the required \ac{pa} back-off are set by this continuous-time peak, we
oversample by $J$ to approximate it closely.} approximates the continuous-time
peak~\cite{6140624}. Finally, a \ac{cp} is
prepended to each symbol to absorb multipath.

At the receiver the operations are inverted: \ac{cp} removal, \ac{fft}, subcarrier
demapping, de-spreading, followed by soft demapping to the per-bit
\acp{llr}~\eqref{eq:llr}, belief-propagation \ac{ldpc} decoding, and, for \ac{ps} and joint
shaping, inversion of the \ac{dm}. The \ac{awgn} channel considered here requires no
equalization.

\subsection{Peak-to-Average Power Ratio}
The \ac{papr} of one \dftsofdm{} symbol, that is
one block of $N$ precoded symbols, is
\begin{equation}
\mathrm{PAPR} \;=\; \frac{\max_{\ell} |s_\ell|^2}{\tfrac{1}{J M_{\mathrm{sc}}}\sum_{\ell=0}^{J M_{\mathrm{sc}}-1}|s_\ell|^2}.
\label{eq:papr}
\end{equation}
\Ac{papr} varies from symbol to
symbol, so we characterize its distribution over many transmitted symbols through the \ac{ccdf}
$\Pr(\mathrm{PAPR}>\xi)$, following standard practice for \ac{ofdm} and single-carrier
waveforms~\cite{rahmatallah2013papr}, and read off the \ac{papr} at a fixed \ac{ccdf} level.

\subsection{Probabilistic Amplitude Shaping}
\label{sec:pas}
We realize \ac{ps} and joint shaping with the \ac{pas} architecture~\cite{7307154}, which
applies the shaped distribution to the \ac{qam} amplitudes through a \ac{dm}, here a
\ac{ccdm}~\cite{7322261}. Of the $m$ bits of a \ac{qam} symbol, the \ac{dm} fills the $m-2$ amplitude
positions, while the two sign positions carry $\gamma_{\mathrm{QAM}}$ uniform information bits and
$2-\gamma_{\mathrm{QAM}}$ parity bits of the systematic \ac{fec} encoder. The information bits
bypass the \ac{dm} and enter the encoder directly, as shown in Fig.~\ref{fig:tx}. For a \ac{fec}
rate $R$, the bypass carries $\gamma_{\mathrm{QAM}} = mR-(m-2)$ bits per symbol, twice the
per-dimension sign fraction of~\cite{7307154}. For 64-QAM at the rate $R=5/6$ used here, $\gamma_{\mathrm{QAM}}=1$, so one sign
position carries an information bit and the other a parity bit.

\Ac{ps} and joint shaping apply the same distribution to the \ac{pam} amplitude $A$ of both
quadrature dimensions, with a uniform sign per dimension and the two dimensions
independent~\cite{7307154}. The amplitude entropy $H(A)$ therefore refers to one real dimension.
Applied to the two dimensions of \ac{qam}, the per-dimension \ac{pas}
rate of~\cite{7307154} gives the \ac{seff}
\begin{equation}
\mathrm{SE} \;=\; 2\,r_{\mathrm{dm}} \;+\; \gamma_{\mathrm{QAM}} \;=\; 2\,r_{\mathrm{dm}} \;+\;
mR-(m-2),
\label{eq:pas-se}
\end{equation}
in \ac{bpcu}, where $r_{\mathrm{dm}}=H(A)-r_{\mathrm{loss}}$ is the rate the \ac{dm}
delivers per real dimension, the amplitude entropy less the rate loss
$r_{\mathrm{loss}}$ of the finite-length \ac{ccdm}. For 64-QAM ($m=6$) the \ac{seff} relation reads
$\mathrm{SE}=2r_{\mathrm{dm}}+6R-4$. More aggressive shaping lowers $r_{\mathrm{dm}}$, so at
fixed \ac{seff} it forces a higher \ac{fec} rate $R$, which only the sign bypass makes possible.

Rearranging the rate bound of~\cite{7322261} upper bounds the
\ac{ccdm} rate loss,
\begin{equation}
r_{\mathrm{loss}} \;\le\; \frac{(|\mathcal{A}|-1)\,\log_2(N+|\mathcal{A}|-1)}{N} + \frac{1}{N},
\label{eq:ccdmrate}
\end{equation}
where $\mathcal{A}$ is the amplitude alphabet. This bound is used as the starting estimate that sets the initial
entropy target for the optimization in Section~\ref{sec:cost}.

\section{NET-GAIN OPTIMIZATION FRAMEWORK}
\label{sec:framework}
We compare all constellations against the uniform, i.e., equiprobable, 64-QAM reference at a matched \ac{seff}. Let
$\mathrm{SNR}^\star$ denote the \ac{snr} at which a given constellation reaches a target
coded \ac{bler} $\epsilon^\star$, and let $\mathrm{PAPR}^\star$ denote the \ac{papr}
at which the \ac{ccdf} drops to a target level $\delta^\star$. The communication gain, the \ac{papr} gain, and
the net-gain, all in decibels, are
\begin{align}
G_{\mathrm{com}} &= \mathrm{SNR}^\star_{\mathrm{unif}} - \mathrm{SNR}^\star_{\mathrm{shaped}},
\label{eq:gcom}\\
G_{\mathrm{PAPR}} &= \mathrm{PAPR}^\star_{\mathrm{unif}} - \mathrm{PAPR}^\star_{\mathrm{shaped}},
\label{eq:gpapr}\\
G_{\mathrm{net}} &= G_{\mathrm{com}} + G_{\mathrm{PAPR}}.
\label{eq:gnet}
\end{align}
By this sign convention, a positive \ac{papr} gain $G_{\mathrm{PAPR}}$ means shaping lowers the
\ac{papr} below the uniform reference, while a negative value means it raises the \ac{papr}. A
constellation is beneficial on the \ac{ul} only when $G_{\mathrm{net}}>0$, that is when the
communication gain it provides exceeds the \ac{papr} it costs. We formulate constellation design as the
maximization of this net-gain, approached through the optimization problem below.

\subsection{Constellation Optimization}
The coded \ac{bler} that enters~\eqref{eq:gcom} cannot be evaluated inside a gradient-based
loop, because it requires a full decoder run per candidate constellation. We therefore optimize
on differentiable surrogates: the \ac{bmd} rate~\eqref{eq:gmi} stands
in for the communication performance, and a moment-based surrogate of the single-carrier
amplitude statistics stands in for the \ac{papr}. With these surrogates the
design solves
\begin{align}
\min_{\mathcal{X},\,P_X} \quad & \mathcal{L}_{\mathrm{PAPR}}(\mathcal{X},P_X)
\label{eq:problem}\\
\text{s.t.} \quad & R_{\mathrm{BMD}}(X) \ge R_{\mathrm{BMD}}^{\min}, \nonumber\\
& H(X) = H_{\mathrm{t}},  \nonumber\\ &\textstyle\sum_i P_X(x_i)\,|x_i|^2 = 1, \nonumber
\end{align}
that is, the design minimizes the \ac{papr} surrogate $\mathcal{L}_{\mathrm{PAPR}}$. The first
constraint places a floor on the \ac{air}. Without it, the optimizer could trivially minimize
\ac{papr} at the expense of communication performance. The second fixes the entropy
$H(X)$ to a target $H_{\mathrm{t}}$, which pins the realized \ac{seff} through the \ac{pas} rate
relation~\eqref{eq:pas-se}. This constraint is not needed for \ac{gs}, whose probabilities stay uniform throughout
the optimization, so its entropy never changes. The third is a unit-average-power constraint,
following the \ac{nuc} design of~\cite{7419222}.

The free
parameters are the constellation point positions, the symbol probabilities, or both, depending on
the \ac{cs} methods. \Ac{gs} optimizes the \ac{2d} point positions at uniform
probabilities, forming a \ac{2d} \ac{nuc} whose full 64-point layout is generated
from the first quadrant by mirroring, so that quadrant symmetry reduces the geometric degrees of
freedom by a factor of four. \Ac{ps} instead
fixes the grid and optimizes a per-dimension \ac{pmf}, the \ac{1d} reduction that keeps
the shaping realizable by \ac{ccdm}~\cite{7322261} alone. Joint shaping optimizes
both, but constrains its geometry to a \ac{1d} \ac{nuc} so that the shaping
stays separable per dimension and can be realized by the same \ac{1d}
\ac{ccdm}~\cite{7322261} as \ac{ps}. The probabilities are learned as unconstrained
logits mapped by the softmax function to the per-dimension \ac{pmf}, which keeps $P_X$ a valid distribution and
makes the shaping differentiable~\cite{Goodfellow-et-al-2016,9024567}.

Because the \ac{air} is a non-convex function of the
constellation~\cite{7419222}, it is optimized numerically rather than in closed form. The design
is a differentiable learning loop: the learnable \ac{pam}
amplitudes and probability logits build the constellation and its \ac{pmf}, from which the
metrics and the cost~\eqref{eq:cost} follow, and backpropagation through this pipeline yields
Adam~\cite{kingma2017adammethodstochasticoptimization} updates to the geometry and probability
parameters in separate steps. As in the per-code-rate \ac{nuc} design
of~\cite{7419222}, where each constellation is optimized at the target \ac{snr} of its \ac{fec}
rate, we fix a single design \ac{snr} shared by all schemes, defined as the \ac{snr} at which
uniform 64-QAM attains the target \ac{seff}, that is the \ac{snr} where the \ac{bmd} rate
of the uniform reference equals \num{4}~\ac{bpcu}.

\subsection{Cost Function}\label{sec:cost}
We solve~\eqref{eq:problem} by turning its constraints into weighted penalty terms and minimizing
the resulting cost
\begin{equation}
\begin{aligned}
\mathcal{L} \;=\;
\mathcal{L}_{\mathrm{PAPR}}
&\;+\; \alpha\,\big[R_{\mathrm{BMD}}^{\min}-\widehat{R}_{\mathrm{BMD}}\big]^{+}\\
&\;+\; \beta\,\big(H(X)-H_{\mathrm{t}}\big)^{2}
\;+\; \mu\,\big(\bar{P}-1\big)^{2},
\end{aligned}
\label{eq:cost}
\end{equation}
where $[\cdot]^{+}=\max(0,\cdot)$ and $\bar{P}=\sum_i P_X(x_i)\,|x_i|^2$ is the average symbol
power. The four terms are, in order, the \ac{papr} cost, a rate-floor penalty, an
entropy-equality constraint, and a power-normalization penalty. The weights $\alpha$,
$\beta$, and $\mu$ set their relative importance. We select $\alpha$, $\beta$, $\mu$, the
rate floor $R_{\mathrm{BMD}}^{\min}$, and the learning rate by a hyperparameter search.

The \ac{papr} cost is realized by the closed-form kurtosis metric, the
normalized fourth moment of the alphabet
\begin{equation}
\mathcal{L}_{\mathrm{PAPR}} \;=\; \kappa \;=\;
\frac{\sum_i P_X(x_i)\,|x_i|^4}{\big(\sum_i P_X(x_i)\,|x_i|^2\big)^2},
\label{eq:kurtosis}
\end{equation}
which is minimized ($\kappa\!\to\!1$) by a constant-envelope alphabet.

The rate term acts on the \ac{bmd} rate~\eqref{eq:gmi}, the \ac{air} protected by the floor,
approximated by the \ac{mc} estimation $\widehat{R}_{\mathrm{BMD}}$~\eqref{eq:gmi-mc}. This
penalty $\alpha[R_{\mathrm{BMD}}^{\min}-\widehat{R}_{\mathrm{BMD}}]^{+}$ applies only when the \ac{bmd} rate drops below the floor
$R_{\mathrm{BMD}}^{\min}$, so the optimizer is free to reduce \ac{papr} as long as the rate target is met.

The entropy constraint pins $H(X)$ to a target $H_{\mathrm{t}}$, essential for a fair comparison
since all schemes operate at the same \ac{seff}, so the amount of shaping is fixed through the
\ac{pas} rate relation~\eqref{eq:pas-se}. Using the
\ac{ccdm} rate-loss bound~\eqref{eq:ccdmrate} of
Section~\ref{sec:system}, the entropy target $H_{\mathrm{t}}$ is chosen so that, after subtracting this loss, the
realized rate~\eqref{eq:pas-se} equals the target \ac{seff}. Because the true \ac{ccdm} loss
depends on the current \ac{pmf}, \ac{ps} and joint shaping additionally apply an
entropy-adaptation step that re-targets $H_{\mathrm{t}}$ to the current \ac{ccdm} loss, updated
periodically during learning. This keeps the effective \ac{seff} constant across all schemes and makes the net-gain comparison fair.
During learning, the bit-to-symbol labeling is likewise periodically reassigned by an
I/Q-aware Gray rule as the constellation points move.

\section{SIMULATION SETUP AND RESULTS}
\label{sec:results}
We now validate the constellations optimized with the framework of
Section~\ref{sec:framework} on a complete \dftsofdm{} link-level simulation, and read the
net-gain from the coded link.
Table~\ref{tab:sim} summarizes the simulation parameters. For \ac{ps} and joint shaping,
information bits are shaped by \ac{ccdm}~\cite{7322261} of block length \num{384} and encoded by the
3GPP 5G \ac{nr} \ac{ldpc} code~\cite{3gpp-38.212} within the \ac{pas} architecture~\cite{7307154}.
\Ac{gs} keeps uniform symbol probabilities, so its information bits are encoded directly by the
\ac{nr} \ac{ldpc} code, without \ac{ccdm} or \ac{pas}. All schemes map their coded bits to a
64-point constellation and transmit the resulting symbols over an \ac{awgn} channel with the
\dftsofdm{} waveform~\cite{myung2006singlecarrier}. The receiver performs soft demapping,
belief-propagation \ac{ldpc} decoding, and, for \ac{ps} and joint shaping, \ac{ccdm} dematching to
recover the information bits. For \ac{ps} and joint shaping, the entropy-adaptation step
of Section~\ref{sec:framework} produces the shaped amplitudes, so that every scheme operates at
the same target \ac{seff}. The communication performance is taken as the \ac{snr} at which the
coded \ac{bler} first drops below $\epsilon^\star=10^{-2}$, and the per-symbol \ac{papr} is read at
$\mathrm{CCDF}=\delta^\star=10^{-3}$.

\Ac{pas} requires that the amplitude bits, shaped by the \ac{dm}, reach the symbol
mapper unmodified. Standard conform \ac{ulsch}
and \ac{pusch} bit processing does not guarantee this. We therefore retain the 5G \ac{nr}
\ac{ldpc} code itself, i.e., its \ac{bg}, lifting, and parity structure, but omit the
rate-matching bit interleaver and the \ac{pusch} scrambler~\cite{3gpp-38.212,3gpp-38.211}, either of
which would remove the effect of \ac{ps}. 5G \ac{nr} \ac{ldpc} encoding never
transmits the leading $2Z_{\mathrm{c}}$ systematic
bits~\cite{3gpp-38.212}. We therefore fill these positions with
uniform sign bits, so that no shaped amplitude bits are punctured, as otherwise the \ac{pas} would be destroyed.

We compare five constellations against the uniform 64-QAM reference. Two are literature baselines,
plain \ac{mb} \ac{ps}~\eqref{eq:mb} and the ATSC~3.0 64-\ac{nuc}~\cite{7419222}. The other three
are the net-gain-optimized \ac{ps}, \ac{gs}, and joint constellations of
Section~\ref{sec:framework}. \Ac{ps} and joint shaping use \ac{pas} at \ac{fec} rate $R=5/6$,
whereas \ac{gs} keeps uniform probabilities at $R=2/3$, both rates chosen so that every scheme
reaches the same target \ac{seff}. Fig.~\ref{fig:consts} shows the three
optimized constellations, Fig.~\ref{fig:papr} the \ac{papr} \acp{ccdf}, and Fig.~\ref{fig:bler}
the coded \ac{bler} curves. Table~\ref{tab:results} then collects the resulting gains of
\eqref{eq:gcom}-\eqref{eq:gnet}.

\begin{table}[t]
\caption{Link-level simulation parameters.}
\label{tab:sim}
\centering
\footnotesize
\begin{tabular}{lc}
\toprule
\textbf{Parameter} & \textbf{Value} \\
\midrule
Modulation                         & 64-QAM \\
Waveform                           & DFT-s-OFDM \\
DFT-precoder size $N$              & 96 (8 PRB) \\
Subcarrier grid $M_{\mathrm{sc}}$  & 256 \\
PAPR oversampling factor $J$       & 8 \\
DFT-s-OFDM symbols per CCDM block & 4 \\
Channel                            & AWGN \\
FEC                                & 5G NR LDPC~\cite{3gpp-38.212} \\
\;\; \ac{bg}                        & 1 (all branches) \\
\;\; $R$, $Z_{\mathrm{c}}$ (PS / Joint)   & $5/6$, 88 \\
\;\; $R$, $Z_{\mathrm{c}}$ (GS / uniform) & $2/3$, 72 \\
Distribution matcher               & CCDM, block length $384$ \\
Target \ac{seff}                & $4$ bpcu \\
\bottomrule
\end{tabular}
\end{table}

\begin{figure}[t]
\centering
\resizebox{\columnwidth}{!}{%
\begin{tikzpicture}
\begin{groupplot}[
  group style={group size=3 by 1, horizontal sep=0.5cm},
  width=2.5cm, height=2.5cm, scale only axis,
  xmin=-1.85, xmax=1.85, ymin=-1.85, ymax=1.85,
  xtick={-1,0,1}, ytick={-1,0,1},
  tick label style={font=\scriptsize}, title style={font=\small},
  grid=both, major grid style={solid, gray!15}, minor tick num=0,
]
\nextgroupplot[title={PS}]
\addplot[scatter, only marks, mark=*, color=clr4,
  point meta=explicit symbolic,
  scatter/@pre marker code/.style={/tikz/mark size=\pgfplotspointmeta},
  scatter/@post marker code/.style={}]
  table[x=x, y=y, meta=ms] {figs/const_ps.dat};
\nextgroupplot[title={GS}]
\addplot[scatter, only marks, mark=*, color=clr5,
  point meta=explicit symbolic,
  scatter/@pre marker code/.style={/tikz/mark size=\pgfplotspointmeta},
  scatter/@post marker code/.style={}]
  table[x=x, y=y, meta=ms] {figs/const_gs.dat};
\nextgroupplot[title={Joint}]
\addplot[scatter, only marks, mark=*, color=clr6,
  point meta=explicit symbolic,
  scatter/@pre marker code/.style={/tikz/mark size=\pgfplotspointmeta},
  scatter/@post marker code/.style={}]
  table[x=x, y=y, meta=ms] {figs/const_joint.dat};
\end{groupplot}
\end{tikzpicture}%
}
\caption{Net-gain-optimized 64-QAM constellations at unit average power, with marker radius
proportional to the symbol probability. \Ac{ps} shapes the probabilities on a uniform grid,
\ac{gs} relocates the constellation points at uniform probability (\ac{2d} \ac{nuc}), and the joint
scheme combines a \ac{1d} \ac{nuc} grid with a shaped per-dimension \ac{pmf}.}
\label{fig:consts}
\end{figure}
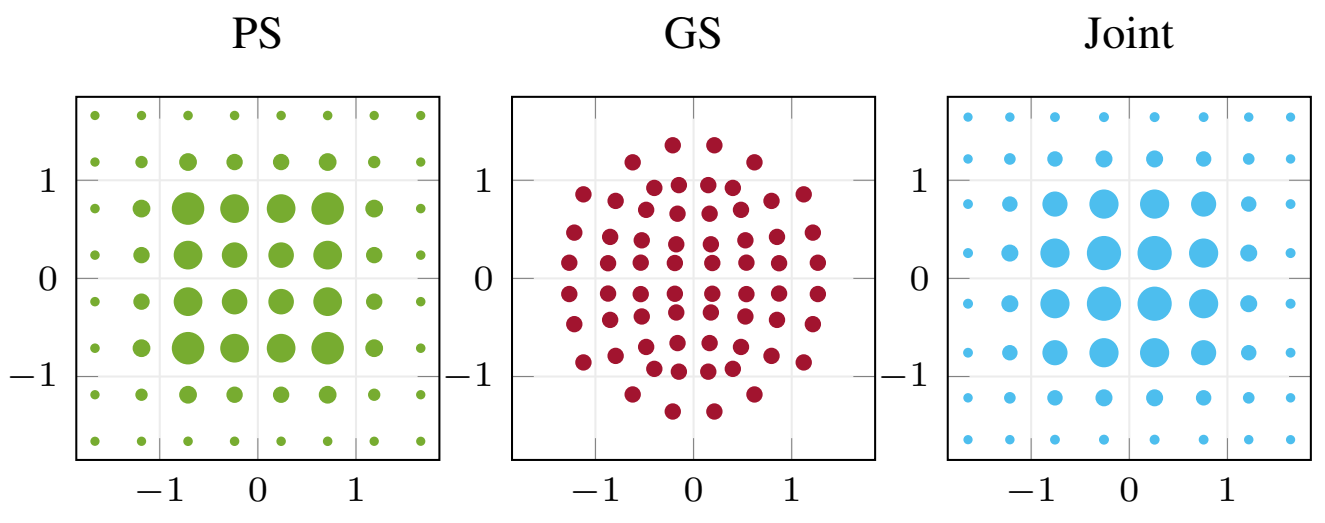

\begin{figure}[t]
\centering
\begin{tikzpicture}
\begin{semilogyaxis}[resultaxis,
  xlabel={Per-symbol PAPR $\xi$ (\si{\deci\bel})},
  ylabel={$\Pr(\mathrm{PAPR}>\xi)$},
  xmin=5, xmax=10.5, ymin=1e-4, ymax=1,
  legend style={fill=white, fill opacity=0.8,text opacity =1,at={(0.02,0.02)}, anchor=south west},
]
\addplot[clr1, mark=o,            mark repeat=3, mark phase=0] table[col sep=comma, x=papr_dB, y=Uniform64QAM] {figs/ccdf_papr.dat}; \addlegendentry{Uniform}
\addplot[clr2, mark=triangle*, mark repeat=3, mark phase=1] table[col sep=comma, x=papr_dB, y=MB_H_5_1_] {figs/ccdf_papr.dat}; \addlegendentry{MB PS}
\addplot[clr3, mark=square*,      mark repeat=3, mark phase=2] table[col sep=comma, x=papr_dB, y=ATSC3_0NUC] {figs/ccdf_papr.dat}; \addlegendentry{ATSC 3.0 NUC}
\addplot[clr4, mark=diamond*,   mark repeat=3, mark phase=0] table[col sep=comma, x=papr_dB, y=PSNetgain] {figs/ccdf_papr.dat}; \addlegendentry{PS opt.}
\addplot[clr5, mark=pentagon*,  mark repeat=3, mark phase=1] table[col sep=comma, x=papr_dB, y=GSNetgain] {figs/ccdf_papr.dat}; \addlegendentry{GS opt.}
\addplot[clr6, mark=asterisk, mark size=2.2pt,  mark repeat=3, mark phase=2] table[col sep=comma, x=papr_dB, y=JointNetgain] {figs/ccdf_papr.dat}; \addlegendentry{Joint opt.}
\draw[red, dashed, line width=1.2pt] (axis cs:5,1e-3) -- (axis cs:10.5,1e-3);
\end{semilogyaxis}
\end{tikzpicture}
\caption{Per-symbol PAPR CCDF on DFT-s-OFDM for the compared constellations. The
horizontal dashed line marks the $\mathrm{CCDF}=10^{-3}$ level at which \ac{papr}$^\star$ is read.}
\label{fig:papr}
\vspace{2mm}
\begin{tikzpicture}
\begin{semilogyaxis}[resultaxis,
  xlabel={SNR (\si{\deci\bel})},
  ylabel={BLER},
  xmin=12.0, xmax=14.5, ymin=1e-3, ymax=1,
  legend style={fill=white, fill opacity=0.8,text opacity =1,at={(0.02,0.02)}, anchor=south west},
]
\addplot[clr1, mark=o,            mark repeat=3, mark phase=0] table[x=snr, y=b0] {figs/bler.dat}; \addlegendentry{Uniform}
\addplot[clr2, mark=triangle*, mark repeat=3, mark phase=1] table[x=snr, y=b1] {figs/bler.dat}; \addlegendentry{MB PS}
\addplot[clr3, mark=square*,      mark repeat=3, mark phase=2] table[x=snr, y=b2] {figs/bler.dat}; \addlegendentry{ATSC 3.0 NUC}
\addplot[clr4, mark=diamond*,   mark repeat=3, mark phase=0] table[x=snr, y=b3] {figs/bler.dat}; \addlegendentry{PS opt.}
\addplot[clr5, mark=pentagon*,  mark repeat=3, mark phase=1] table[x=snr, y=b4] {figs/bler.dat}; \addlegendentry{GS opt.}
\addplot[clr6, mark=asterisk, mark size=2.2pt, mark repeat=3, mark phase=2] table[x=snr, y=b5] {figs/bler.dat}; \addlegendentry{Joint opt.}
\draw[red, dashed, line width=1.2pt] (axis cs:12,1e-2) -- (axis cs:14.5,1e-2);
\end{semilogyaxis}
\end{tikzpicture}
\caption{Coded \ac{bler} versus SNR. The horizontal dashed line marks
the $\mathrm{BLER}=10^{-2}$ point at which $\mathrm{SNR}^\star$ is read.}
\label{fig:bler}
\end{figure}

\begin{table}[t]
\caption{Communication gain, PAPR gain, and net-gain relative to uniform 64-QAM.}
\label{tab:results}
\centering
\begin{tabular}{lccc}
\toprule
\textbf{Constellation} & $G_{\mathrm{com}}$ (\si{\deci\bel}) & $G_{\mathrm{PAPR}}$ (\si{\deci\bel}) & $\mathbf{G_{\mathrm{net}}}$ (\si{\deci\bel}) \\
\midrule
Uniform 64-QAM (ref.)        & $-$ & $-$ & $-$ \\
\ac{mb} PS        & $+0.98$ & $-1.26$ & $-0.28$ \\
ATSC 3.0 NUC (GS)            & $+0.40$ & $-0.26$ & $+0.14$ \\
PS, net-gain-optimized       & $+0.49$ & $-0.46$ & $+0.03$ \\
\textbf{GS, net-gain-optimized}       & $\mathbf{+0.28}$ & $\mathbf{+0.18}$ & $\mathbf{+0.46}$ \\
Joint, net-gain-optimized    & $+0.79$ & $-0.69$ & $+0.10$ \\
\bottomrule
\end{tabular}
\end{table}

The results show that the two literature baselines already expose the trade-off the framework
targets. With \SI{0.98}{\deci\bel} conventional \ac{mb} \ac{ps} earns the largest communication gain of any constellation in
Table~\ref{tab:results}, but raises the \ac{papr} by an even larger
\SI{1.26}{\deci\bel}, leaving a negative net-gain of \SI{-0.28}{\deci\bel}. The rate improvement
that motivates shaping is thus consumed entirely by the \ac{papr} penalty. The ATSC~3.0 \ac{nuc},
a fixed geometric design, shows the reverse behaviour, converting a more modest communication gain
into a positive net-gain of \SI{+0.14}{\deci\bel} because its \ac{papr} cost stays small.

Net-gain-aware optimization improves every \ac{cs} method over its own baseline, though only
\ac{gs} reaches a clear gain. Within \ac{ps}, the optimized
\ac{ps} moves from the \SI{-0.28}{\deci\bel}
of \ac{mb} to \SI{+0.03}{\deci\bel}, trading part of the communication gain for a smaller \ac{papr}
increase of \SI{0.46}{\deci\bel}, a swing of \SI{0.31}{\deci\bel} that removes the deficit but
leaves the scheme almost at break-even. The joint scheme, whose geometry is a \ac{1d} \ac{nuc}, reaches a net-gain of \SI{+0.10}{\deci\bel}. Within
\ac{gs}, the optimized \ac{2d} \ac{nuc} improves on the ATSC~3.0 \ac{nuc} from
\SI{+0.14}{\deci\bel} to \SI{+0.46}{\deci\bel}, the largest net-gain of all. It is also the only
scheme that improves both partial gains at once, lowering the \ac{papr} below the uniform reference
for a positive \ac{papr} gain of \SI{+0.18}{\deci\bel} while still providing a
\SI{0.28}{\deci\bel} communication gain.

The robustness of \ac{gs} follows from the degrees of freedom of its \ac{2d} \ac{nuc}. Relocating
the constellation points equalizes the symbol amplitudes, which is what lowers the \ac{papr} instead
of raising it. Because its probabilities stay uniform it needs no \ac{dm} and therefore pays no
rate loss. \Ac{ps} and joint shaping, in contrast, realize their shaped distribution through a
\ac{dm}. At the finite block length used here, the \ac{ccdm}
rate loss~\eqref{eq:ccdmrate} is non-negligible and must be recovered through a higher \ac{fec}
rate to still reach the fixed target \ac{seff}, which weakens the code and reduces the communication gain. \Ac{gs} is also the simplest to
implement, needing only a soft demapper matched to the \ac{nuc} point positions.

\section{CONCLUSION}
\label{sec:conclusion}
We framed \ac{cs} for the \dftsofdm{} 5G \ac{nr} \ac{ul} as the maximization of a single
net-gain metric, the sum of the coded communication gain and the \ac{papr} gain, turning the
trade-off characterized in~\cite{8227810} into an explicit design objective. On a complete
\dftsofdm{} and 5G \ac{nr} \ac{ldpc} link-level simulation setup, conventional \ac{mb} \ac{ps} yields
a negative net-gain, since its \ac{papr} increase exceeds its communication gain, whereas
net-gain-aware shaping removes this deficit for all \ac{cs} methods. However, only \ac{gs} turns it
into a substantial gain, while the optimized \ac{ps} reaches no more than break-even. \Ac{gs} is the
most robust choice, even
reducing \ac{papr} below the uniform reference, which it also does at 16- and 256-QAM.
Future work includes evaluation over fading channels and validating the optimized
constellations against a realistic \ac{pa} model.

\bibliographystyle{IEEEtran}
\bibliography{refs}

\end{document}

%% file: acronyms.tex
\begin{acronym}[MPDM]
	\acro{jcas}[JCAS]{joint communication and sensing}
	\acro{jrc}[JRC]{joint radar-communications}
	\acro{ue}[UE]{user equipment}
	\acro{nr}[NR]{New Radio}
	\acro{3gpp}[3GPP]{3rd Generation Partnership Project}
	\acro{ofdm}[OFDM]{orthogonal frequency division multiplexing}
	\acro{ofdma}[OFDMA]{orthogonal frequency division multiple access}
	\acro{sc-fdma}[SC-FDMA]{single-carrier frequency division multiple access}
	\acro{ul}[UL]{uplink}
	\acro{dl}[DL]{downlink}
	\acro{srs}[SRS]{sounding reference signal}
	\acro{bs}[BS]{base station}
	\acro{fft}[FFT]{fast Fourier transform}
	\acro{papr}[PAPR]{peak-to-average power ratio}
	\acro{cazac}[CAZAC]{constant-amplitude zero-autocorrelation}
	\acro{zc}[ZC]{Zadoff-Chu}
	\acro{rf}[RF]{radio frequency}
	\acro{czt}[CZT]{chirp-Z transform}
	\acro{fpga}[FPGA]{Field Programmable Gate Array}
	\acro{fr2}[FR2]{frequency range 2}
	\acro{qam}[QAM]{quadrature amplitude modulation}
	\acro{pam}[PAM]{pulse-amplitude modulation}
	\acro{qpsk}[QPSK]{quadrature phase shift keying}
	\acro{bpsk}[BPSK]{binary phase shift keying}
	\acro{sll}[SLL]{sidelobe level}
	\acro{psl}[PSL]{peak sidelobe level}
	\acro{isl}[ISL]{integrated sidelobe level}
	\acro{dft}[DFT]{discrete Fourier transform}
	\acro{tdd}[TDD]{time division duplex}
	\acro{rb}[RB]{resource block}
	\acro{prb}[PRB]{physical resource block}
	\acro{dd}[DD]{delay-Doppler}
	\acro{otfs}[OTFS]{Orthogonal Time Frequency Space}
	\acro{sfft}[SFFT]{symplectic finite fourier transform}
	\acro{isfft}[ISFFT]{inverse symplectic finite fourier transform}
	\acro{ifft}[IFFT]{inverse fast Fourier transform}
	\acro{fir}[FIR]{finite impulse response}
	\acro{mimo}[MIMO]{multiple-input multiple-output}
	\acro{tf}[TF]{time-frequency}
	\acro{snr}[SNR]{signal-to-noise ratio}
	\acro{ccdf}[CCDF]{complementary cumulative distribution function}
	\acro{tx}[TX]{transmitter}
	\acro{rx}[RX]{receiver}
	\acro{mf}[MF]{matched-filter}
	\acro{rmse}[RMSE]{root mean square error}
	\acro{adc}[ADC]{analog-to-digital converter}
	\acro{dac}[DAC]{digital-to-analog converter}
	\acro{crlb}[CRLB]{Cram\'er-Rao lower bound}
	\acro{fim}[FIM]{Fisher information matrix}
	\acro{sccd}[SCCD]{sparse cyclic coordinate descent}
	\acro{socp}[SOCP]{second-order cone programming}
	\acro{awgn}[AWGN]{additive white Gaussian noise}
	\acro{fdss}[FDSS]{frequency-domain spectral shaping}
	\acro{cp}[CP]{cyclic prefix}
	\acro{evm}[EVM]{error vector magnitude}
	\acro{ber}[BER]{bit error rate}
	\acro{mrc}[MRC]{maximum-ratio combining}
	\acro{zf}[ZF]{zero-forcing}
	\acro{pa}[PA]{power amplifier}
	\acro{obo}[OBO]{output back-off}
	\acro{ibo}[IBO]{input back-off}
	\acro{oobe}[OOBE]{out-of-band emission}
	\acro{aclr}[ACLR]{adjacent channel leakage ratio}
	\acro{pusch}[PUSCH]{physical uplink shared channel}
	\acro{lte}[LTE]{long term evolution}
	\acro{ml}[ML]{machine learning}
	\acro{scs}[SCS]{splitting conic solver}
	\acro{rrc}[RRC]{root-raised cosine}
	\acro{ibe}[IBE]{in-band emission}
	\acro{obw}[OBW]{occupied bandwidth}
	\acro{dmrs}[DM-RS]{demodulation reference signal}
	\acro{idft}[IDFT]{inverse discrete Fourier transform}
	\acro{tdl}[TDL]{tapped delay line}
	\acro{sc}[SC]{subcarrier}
	\acro{ps}[PS]{probabilistic shaping}
	\acro{gs}[GS]{geometric shaping}
	\acro{pas}[PAS]{probabilistic amplitude shaping}
	\acro{dm}[DM]{distribution matcher}
	\acro{ccdm}[CCDM]{constant-composition distribution matching}
	\acro{mpdm}[MPDM]{multiset-partition distribution matching}
	\acro{air}[AIR]{achievable information rate}
	\acro{gmi}[GMI]{generalized mutual information}
	\acro{mi}[MI]{mutual information}
	\acro{mc}[MC]{Monte-Carlo}
	\acro{llr}[LLR]{log-likelihood ratio}
	\acro{ldpc}[LDPC]{low-density parity-check}
	\acro{bg}[BG]{base graph}
	\acro{fec}[FEC]{forward error correction}
	\acro{ulsch}[UL-SCH]{uplink shared channel}
	\acro{nuc}[NUC]{non-uniform constellation}
	\acro{1d}[1-D]{one-dimensional}
	\acro{2d}[2-D]{two-dimensional}
	\acro{bmd}[BMD]{bit-metric decoding}
	\acro{bicm}[BICM]{bit-interleaved coded modulation}
	\acro{bler}[BLER]{block error rate}
	\acro{pmf}[PMF]{probability mass function}
	\acro{seff}[SE]{spectral efficiency}
	\acro{bpcu}[bpcu]{bits per channel use}
	\acro{isac}[ISAC]{integrated sensing and communication}
	\acro{mb}[MB]{Maxwell--Boltzmann}
	\acro{cs}[CS]{constellation shaping}
\end{acronym}